\documentclass[sigconf,screen]{acmart} % Two-column style - camera ready version

\usepackage{subcaption}

\AtBeginDocument{%
  }

\copyrightyear{2026}
\acmYear{2026}
\setcopyright{cc}
\setcctype{by}
\acmConference[AutomotiveUI '26]{18th International Conference on Automotive User Interfaces and Interactive Vehicular Applications}{September 20--23, 2026}{Gothenburg, Sweden}
\acmBooktitle{18th International Conference on Automotive User Interfaces and Interactive Vehicular Applications (AutomotiveUI '26), September 20--23, 2026, Gothenburg, Sweden}
\acmDOI{10.1145/3828157.3828791}
\acmISBN{979-8-4007-2814-3/2026/09}

\begin{document}

\title{Speak to the City: Multimodal Resolution for Outside-the-Vehicle References}

\author{Alireza Parchami}
\orcid{0000-0003-4727-9750}
\affiliation{
  \institution{Mercedes-Benz Tech Innovation GmbH}
  \city{Ulm}
  \country{Germany}
}
\affiliation{
  \institution{Saarland University, Saarland Informatics Campus}
  \city{Saarbrücken}
  \country{Germany}
}
\email{alireza.parchami.cs@gmail.com}

\author{Artin Saberpour}
\orcid{}
\affiliation{
  \institution{Saarland University, Saarland Informatics Campus}
  \city{Saarbrücken}
  \country{Germany}
}
\email{saberpour@cs.uni-saarland.de}

\author{Robin Connor Schramm}
\orcid{0000-0002-4775-4219}
\affiliation{
  \institution{Mercedes-Benz Tech Innovation GmbH}
  \city{Ulm}
  \country{Germany}
}
\affiliation{
  \institution{RheinMain University of Applied Sciences}
  \city{Wiesbaden}
  \country{Germany}
}
\email{robin.schramm@mercedes-benz.com}

\author{Jürgen Steimle}
\orcid{}
\affiliation{
  \institution{Saarland University, Saarland Informatics Campus}
  \city{Saarbrücken}
  \country{Germany}
}
\email{steimle@cs.uni-saarland.de}

\author{Ulrich Schwanecke}
\orcid{0000-0002-0093-3922}
\affiliation{
  \institution{RheinMain University of Applied Sciences}
  \city{Wiesbaden}
  \country{Germany}
}
\email{ulrich.schwanecke@hs-rm.de}

\renewcommand{\shortauthors}{Parchami et al.}

% written by Alireza, last modification 2026.3.14

\begin{abstract}
    As autonomous vehicles and Extended Reality (XR) headsets enable novel in-car interactions, seamlessly querying physical landmarks, known as Outside-the-Vehicle Referencing (OVR), remains challenging due to ego-motion and referential ambiguity. We present a robust, multimodal OVR framework fusing user gaze and natural language to identify Points of Interest (POIs). To address the scarcity of dynamic vehicular data, we developed a VR-based pipeline synchronizing 360-degree transit videos with vehicle GNSS telemetry. Through a user study (N=46) mapping passenger head orientation into a 3D geospatial Digital Twin, we captured authentic gaze-speech behaviors. We subsequently trained a lightweight Transformer network, leveraging LLMs to dynamically align continuous spatial gaze vectors with discrete verbal context. Experimental results demonstrate high accuracy and low computational overhead, achieving an 83.33\% Top-1 accuracy (87.72\% Top-2) and an average inference time of 24.3 milliseconds. This real-time paradigm effectively resolves referential ambiguity, enabling context-aware spatial retrieval for passengers within the vehicle.
\end{abstract}


\begin{teaserfigure}
  \includegraphics[width=\textwidth]{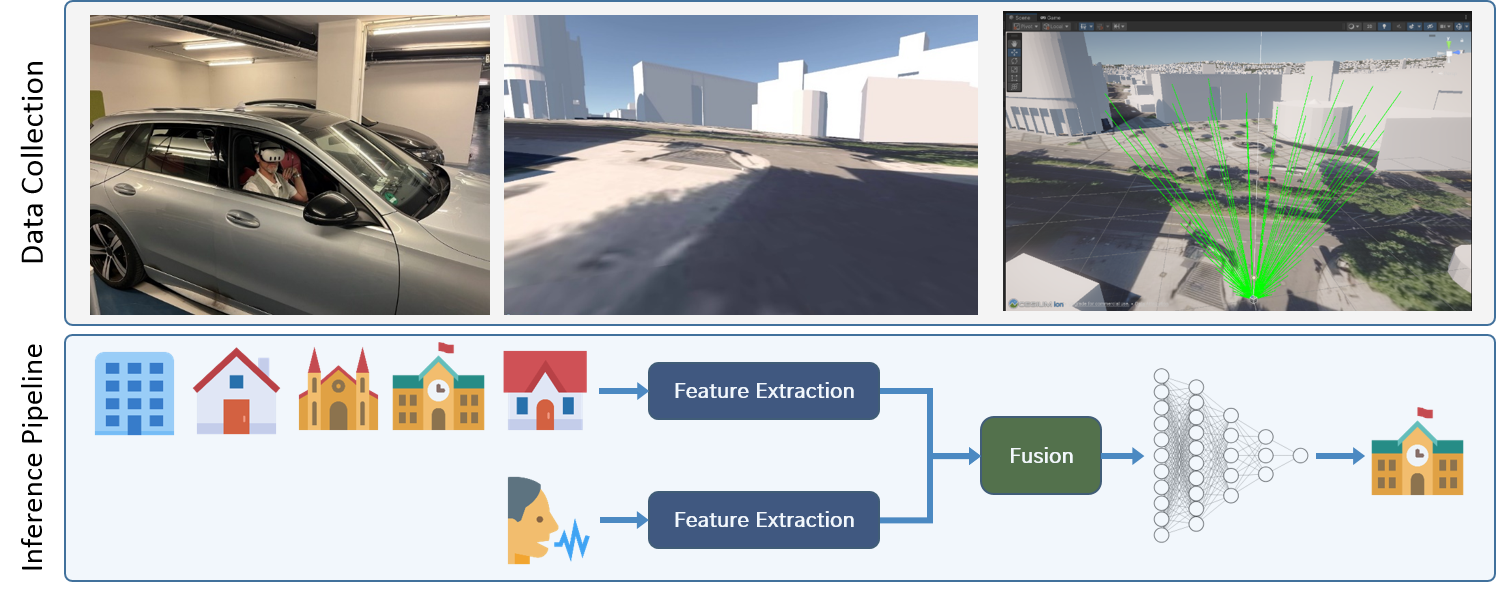}
  \caption{End-to-end overview of our proposed Outside-the-Vehicle Referencing (OVR) framework. \textbf{Top:} The spatial mapping pipeline. In-vehicle passenger head kinematics (left) are projected into a synchronized 3D Digital Twin (middle) to identify candidate buildings via continuous ray-casting (right). \textbf{Bottom:} The multimodal inference architecture, which fuses spatial features from the intersected buildings with discrete semantic speech features via a Transformer network to predict the final target Point of Interest (POI).}
  \Description{}
  \label{fig:teaser}
\end{teaserfigure}

% \received{09 April 2026}
% \received[revised]{09 July 2026}
% \received[accepted]{13 July 2026}

\maketitle

\section{Introduction}
\label{sec:introduction}

The modern automotive landscape is navigating a profound transformation. As vehicles progress toward higher levels of autonomy (SAE Levels 3-5~\cite{J3016_202104}), the paradigmatic definition of the car is shifting from a mere mode of transport into a ``third living space,'' distinct from the home and office. This evolution drives the \textit{Passenger Economy}~\cite{intel2017passenger}, where the primary role of occupants transitions from vehicle operators to active consumers of information and entertainment. Relieved of the cognitive load of driving, passengers are free to engage with their surroundings, naturally tracking landmarks, identifying businesses, and seeking contextual information about the urban environment passing outside their windows~\cite{ardito_designing_2021}.

While modern In-Vehicle Infotainment (IVI) systems offer sophisticated digital interfaces, they remain predominantly introspective—focusing on media, climate control, or 2D route navigation. This creates a significant disconnect between the digital interior of the car and the rich, physical reality outside. In-Car Extended Reality (XR), utilizing Augmented Reality (AR) or Mixed Reality (MR) headsets, presents a profound opportunity to bridge this gap by overlaying digital context directly onto the physical world. However, enabling seamless interaction with the external environment requires these systems to reliably execute \textbf{Outside-the-Vehicle Referencing (OVR)}: the ability to accurately determine a passenger's specific target Point of Interest (POI) within a dense, dynamic urban environment.

Achieving reliable OVR in a moving vehicle is a non-trivial challenge that introduces four core complexities far exceeding those found in static Human-Computer Interaction (HCI):
\begin{enumerate}
    \item \textbf{Environmental Instability:} Passengers are subject to continuous six-degrees-of-freedom (6-DoF) motion and rapidly changing visual scenes, requiring robust transformations between the moving ego-vehicle and geodetically fixed targets.
    \item \textbf{Interaction Ambiguity:} Gaze-based interaction inherently suffers from the \textit{Midas Touch} problem, distinguishing casual scanning from intentional selection~\cite{jacob1990midas}. In dense architectures, relying solely on a head or eye vector lacks the precision to disambiguate true intent among clustered targets. This complexity is also magnified by continuous ego-motion of the vehicle in this context.
    \item \textbf{Data Scarcity:} Collecting labeled ground-truth data in real-world traffic is difficult. Accurately determining which specific building a passenger was looking at during a high-speed drive requires pixel-wise synchronization between world coordinates, vehicle odometry, and human biometrics.
    \item \textbf{Temporal Decay:} Human multimodal interaction is asynchronous. A user might spot a building at time $t$, but issue a verbal query at time $t + \delta$. In a moving vehicle, determining the proper temporal window to match speech with spatial attention is critical for accuracy.
\end{enumerate}

To address these gaps, this paper presents a novel Multimodal Artificial Intelligence (AI) Framework designed to enable frictionless, real-time spatial interaction for passengers in motion. Moving beyond prior OVR research which is predominantly restricted to stationary setups, we specifically target the complexities of OVR in visually dynamic transit environments. Our core hypothesis is that fusing continuous spatial data (head kinematics ray-cast against physical buildings) with discrete semantic data (verbal commands) resolves referential ambiguity. By employing a Transformer-based architecture~\cite{vaswani2017attention} with Cross-Attention mechanisms, our system learns the complex temporal alignment between what a passenger looks at and what they say.

To overcome data scarcity safely, we pioneer a \textbf{Digital Twin Methodology}. We captured synchronized 360-degree urban transit videos and vehicle GNSS/INS telemetry, presenting them to users ($N=46$) in a Virtual Reality (VR) study to capture authentic in-car gaze and speech behavior. By reconstructing the urban environment using Cesium~\cite{CesiumUnity} and OpenStreetMap~\cite{openstreetmap}, we projected passenger head kinematics into the Digital Twin, generating precise ground-truth labels for multimodal neural network training.

\textbf{Contributions:} The primary contributions of this work are three-fold:
\begin{itemize}
    \item A novel, safe, and highly accurate data acquisition pipeline that synchronizes real-world vehicular telemetry with user biometrics inside a 3D geospatial Digital Twin, overcoming the OVR data scarcity bottleneck.
    \begin{sloppypar} % withouth sloppypar, the line exceed col width! this is to make the paragraph with relaxed spacing.
    \item The design and implementation of a lightweight, multimodal Transformer architecture that leverages Sentence-BERT (SBERT) \cite{reimers2019sbert} and Cross-Attention to dynamically fuse spatial gaze candidates with verbal intent.
    \end{sloppypar}
    \item Empirical validation demonstrating that our multimodal fusion strategy successfully mitigates interaction ambiguity and temporal decay during simulated vehicular motion, achieving a Rank-1 prediction accuracy of 83.33\% and outperforming unimodal baseline.
\end{itemize}
\section{Related Work}
\label{sec:related_work}
The convergence of automotive engineering and spatial computing is transforming the vehicle interior into a dynamic digital environment. However, enabling robust OVR requires overcoming significant physical and algorithmic hurdles. We review the foundational literature across In-Car XR and object referencing, highlighting the critical gaps our multimodal framework addresses.

\subsection{In-Car Extended Reality (XR)}
Recent studies have validated the utility of XR headsets for passenger infotainment. Schramm et al. \cite{schramm_blending_2025} deployed a pass-through Head-Mounted Display (HMD) in a moving vehicle to evaluate the visualization of world-fixed POIs. Their field study established baseline parameters for in-car AR (e.g., render distance and information density), demonstrating high user acceptance for autonomous driving scenarios. In a follow-up study \cite{schramm_augmented_2025}, the authors investigated seat-fixed UI visualizations, finding that while users preferred list-based interfaces, mid-air gesture interactions were severely degraded by vehicle motion. They concluded that future in-car XR systems must prioritize robust, hardware-assisted, or multimodal inputs (such as voice) to mitigate the effects of vehicular dynamics. While these works successfully validate the visualization of pre-determined POIs, they do not address the challenge of \textit{input}, allowing passengers to intuitively query unknown buildings in real-time.

\subsection{Inside-the-Vehicle Referencing}
Efforts to enable spatial interaction within the cabin have demonstrated the necessity of multimodal fusion. Aftab et al. \cite{aftab_you_2020} investigated driver attention across 12 interior Regions of Interest (ROIs). By fusing head pose, eye gaze, and finger pointing via deep learning, they achieved 73.7\% accuracy, significantly outperforming unimodal approaches. Their work proved that multimodal systems are highly resilient to individual sensor occlusion. However, camera-based tracking systems restricted to the dashboard often suffer from self-occlusion and a limited field of view (FoV), breaking interaction when users reach beyond the sensor's range. Furthermore, their experiments were conducted in a stationary simulator, completely neglecting the inertial forces and vibrations of a physical vehicle.

\subsection{Outside-the-Vehicle Referencing and Dynamic Challenges}
Extending referencing to the external environment introduces severe complexities. Aftab et al. \cite{aftab_multimodal_2021} applied their multimodal fusion approach to OVR, achieving 72.2\% accuracy when combining head, eye, and finger tracking to select among five external buildings. Similarly, Gomaa et al. \cite{gomaa_looking_2024} utilized a simulated driving environment to fuse speech, head pose, and gaze, introducing an incremental learning algorithm to adapt to individual driver behaviors.

However, relying solely on geometric gaze vectors in dense urban environments introduces the \textit{Midas Touch} problem, the inability to distinguish casual scanning from intentional selection \cite{jacob1990midas}. A single gaze ray might intersect a café, a residential apartment, and a bank simultaneously. Traditional geometric approaches lack the semantic awareness to disambiguate these targets \cite{misu_situated_2014}. 

Furthermore, a critical limitation pervades the current OVR literature: the reliance on stationary setups or low-fidelity driving simulators \cite{aftab_multimodal_2021, gomaa_looking_2024}. Real-world vehicle motion introduces low-frequency inertial forces that disrupt HMD Visual-Inertial Odometry (VIO) and high-frequency vibrations that degrade optical eye-tracking. Additionally, the temporal decay of human interaction, where a passenger looks at a building at time $t$ but speaks at time $t+\delta$, makes reference understanding highly sensitive to timing while moving at speed \cite{misu_situated_2014}. 

\textbf{Bridging the Gap:} Our work addresses these critical limitations by proposing a multimodal architecture that fuses the omnidirectional gaze freedom enabled by XR headsets with the semantic disambiguation power of natural speech. Unlike prior simulator-based studies, we utilize a Digital Twin methodology synchronized with recorded real-world GNSS/INS vehicular telemetry, directly tackling the spatial and temporal complexities of OVR in continuous ego-motion scenarios.
\section{System Overview}

The primary objective of our OVR framework is to enable passengers engaged in urban transit to freely explore and query their urban environment using natural gaze and speech. To achieve this safely and robustly without imposing rigid interaction constraints, we designed an end-to-end, data-driven pipeline (see Figure 1) divided into two primary phases: high-fidelity data acquisition via a Digital Twin, and multimodal POI estimation.

\textbf{Data Acquisition and Digital Twin:} Collecting reliable ground-truth interaction data in live traffic presents profound logistical hurdles. Maintaining stable sensor calibration and environmental consistency across dozens of live commutes introduces variable noise that degrades the integrity of the training data. Moreover, the immense time requirements severely bottleneck the volume of data that can be collected. To overcome these, we developed a VR data collection pipeline utilizing real-world 360-degree urban transit videos strictly synchronized with vehicle state telemetry (GNSS and IMU data). This allowed us to conduct a user study ($N=46$) where participants safely experienced a high-fidelity simulated transit environment while we captured their head kinematics and verbal commands. By reconstructing this physical environment as a 3D Digital Twin using Cesium and OpenStreetMap, we performed continuous ray-casting from the user's perspective to identify the exact buildings intersecting their gaze, generating a robust ground-truth dataset of spatial and contextual metadata.

\textbf{Multimodal Estimation and Aggregation:} To estimate the target POI, we architected a multimodal neural network that processes both visual behavior (gaze sequences) and audio interaction (transcribed speech). The network encodes semantic intent using Sentence-BERT (SBERT) and extracts spatial features from the ray-cast candidate buildings. These modalities are processed via Self-Attention mechanisms and fused using Cross-Attention, effectively weighing geometric gaze data against spoken context. Finally, because a passenger's gaze may continuously intersect same buildings across multiple frames during vehicle motion, the network applies temporal sum aggregation. By accumulating the predicted probabilities for each unique building ID across the interaction window, the system mitigates temporal noise and outputs a highly robust, ranked POI prediction.
\section{Data Acquisition and Digital Twin}
\label{sec:data_acquisition}

A fundamental bottleneck in Outside-the-Vehicle Referencing (OVR) research is the scarcity of continuous-flow, real-world urban datasets. To train a robust multimodal network, we engineered a multi-stage data acquisition pipeline: capturing synchronized in-car 360-degree video, conducting a Virtual Reality (VR) user study to gather behavioral data, and reconstructing the environment via a spatial Digital Twin to generate precise ground-truth labels.

\subsection{Apparatus and Urban Transit Capture}
% We first recorded high-resolution (5.7K, 30 fps) 360-degree videos across three diverse urban routes in Stuttgart and Sindelfingen, Germany. 
We first recorded high-resolution (5.7K, 30 fps) 360-degree videos across three diverse urban routes in Stuttgart (including the Bad Cannstatt district and central areas near Königstraße) and Sindelfingen, Germany. These routes traversed high-density city centers and mixed-use districts, presenting a rich variety of target POIs, including commercial shops, restaurants, and residential buildings, to capture realistic levels of urban density and referential ambiguity. The vehicle operated under normal traffic conditions with speeds mostly ranging from 20 to 45 km/h, encompassing standard urban stops (e.g., traffic lights) and frequent turns across city intersections.

\begin{figure}[h]
    \centering
    \begin{subfigure}[b]{0.2\textwidth}
        \centering
        \includegraphics[width=\textwidth]{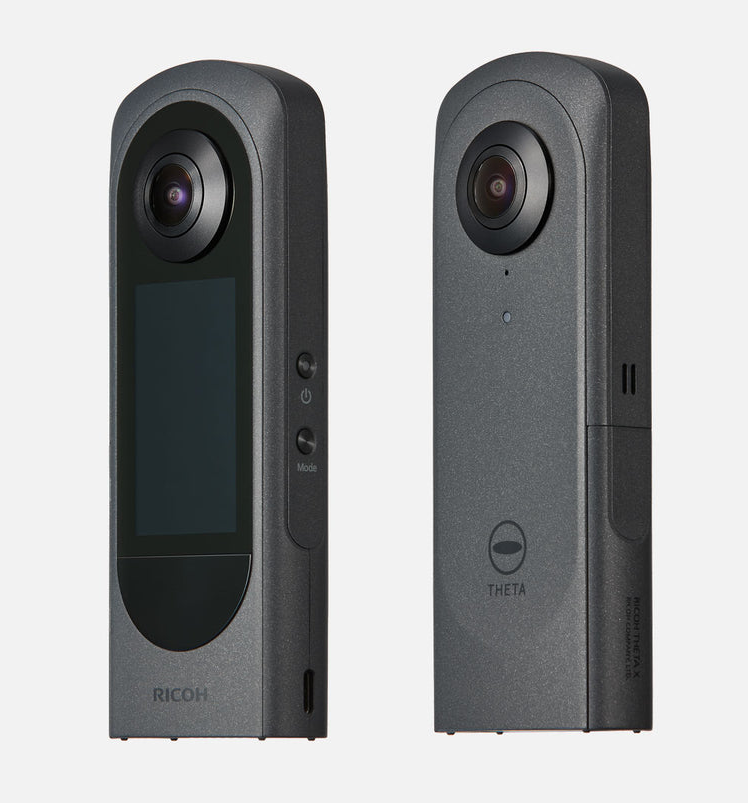}
        \caption{Ricoh Theta X}
    \end{subfigure}
    \hspace{2em}
    \begin{subfigure}[b]{0.2\textwidth}
        \centering
        \includegraphics[width=\textwidth]{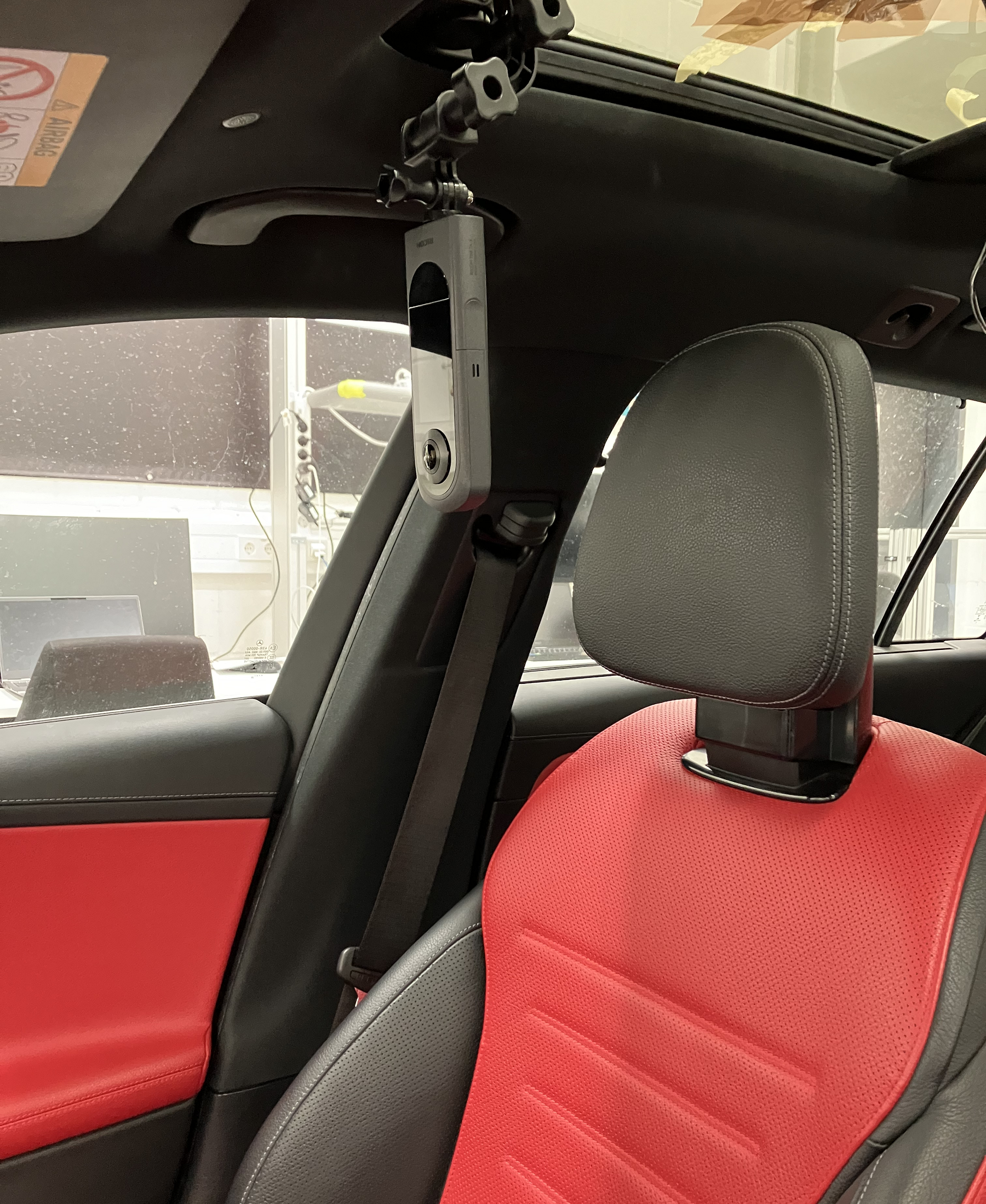}
        \caption{Mounting Position}
    \end{subfigure}
    \caption{Data acquisition hardware setup inside the physical vehicle.} 
    \label{fig:setup_overview}
\end{figure}

The dataset comprises 24 minutes of total drive time (split into one 12-minute and two 6-minute segments), captured entirely during clear daylight conditions around solar noon to ensure consistent visibility. A Ricoh Theta X camera was rigidly mounted on the front passenger seat to approximate a natural passenger's point of view (figure \ref{fig:setup_overview}). 

To enable exact geospatial reconstruction, we utilized custom data logging software to synchronously trigger the video recording and the vehicle's internal telemetry. This guaranteed pixel-perfect synchronization between the video frames and the vehicle's Global Navigation Satellite System (GNSS) and Inertial Measurement Unit (IMU) data.

\subsection{Continuous Ego-Motion VR User Study}
To ensure environmental consistency and robust sensor calibration, we conducted a controlled VR user study ($N=46$, 31Male/15Female, ages 20--60) inside a stationary mid-size sedan. Prior VR/AR experience was measured on a 1--5 scale: the majority were novices (26 at Level 1), with 16 reporting moderate experience (Levels 2--3), and 4 being highly experienced (Levels 4--5). This setup provided realistic physical constraints and passive haptic feedback while participants wore a Meta Quest 3 HMD. 

Participants were initially introduced to the process via an informational task sheet (provided in Appendix \ref{sec:appendixA}) and a brief practice round. Once the study commenced, they experienced the synchronized 360-degree transit videos and naturally queried passing attractions using a push-to-talk controller button. The system continuously logged vehicle GNSS, exact HMD orientation (FoV), and audio. To establish precise ground-truth labels, participants subsequently reviewed their recorded Point-of-View on a monitor and explicitly identified their target OpenStreetMap (OSM) buildings. To mitigate potential recall bias during this retrospective self-identification, participants were provided with their video perspective recording, the spatial vicinity narrowed down on OSM, and playback of their own audio commands in each query to assist in precisely recalling their intended target. A comprehensive breakdown of the 75-minute study procedure, including onboarding and questionnaire timing, is detailed in Appendix \ref{sec:appendixB}.

\subsection{Digital Twin and Ray-Casting Strategy}
\begin{figure}[h]
    \centering
    \begin{subfigure}[b]{0.4\textwidth}
        \centering
        \includegraphics[width=\textwidth]{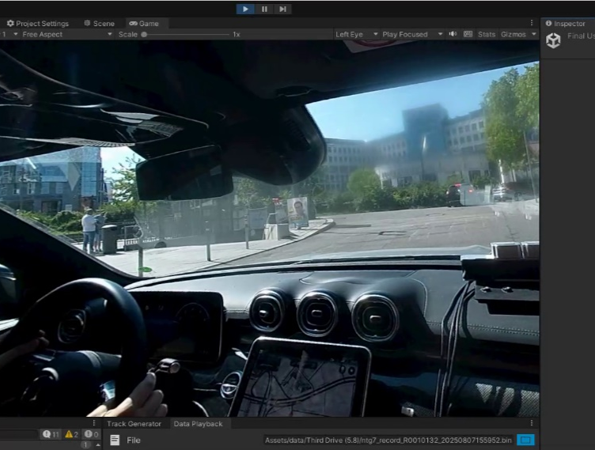}
        \caption{Participant's Point of View in VR study}
    \end{subfigure}
    \hspace{2em}
    \begin{subfigure}[b]{0.4\textwidth}
        \centering
        \includegraphics[width=\textwidth]{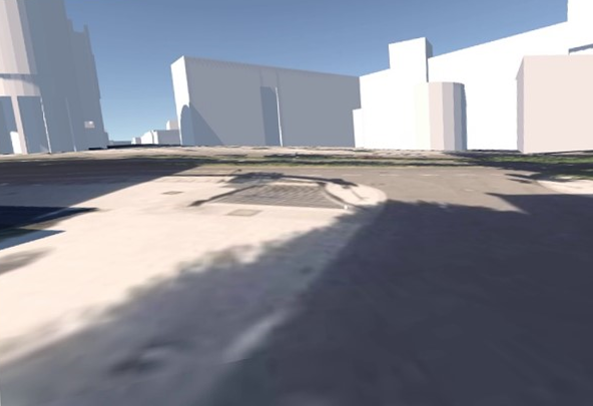}
        \caption{Corresponding Digital Twin Reconstruction}
    \end{subfigure}
    \caption{Spatial correspondence between the visual stimulus and our Digital Twin approach. (a) Participant's Field of View (FoV) during the 360-degree VR transit video. (b) A 2D viewport rendering of the underlying 3D reconstructed scene. Logged vehicle GNSS dictates the Unity virtual camera's global position within Cesium, while logged HMD data (representing the participant's head pose) drives its orientation. This precise alignment replicates the user's view within the 3D space for accurate ray-casting and semantic building extraction.}
    \label{fig:digitalTwin}
\end{figure}

To extract the geometric and semantic features of the buildings passengers looked at, we reconstructed the drive in a 3D Digital Twin utilizing Unity, the Cesium geospatial platform, and OSM building geometry. By injecting the recorded GNSS trajectory and HMD orientation into the Digital Twin, we perfectly recreated the user's physical FoV (Figure \ref{fig:digitalTwin}).

\begin{figure}[h]
    \centering
    \begin{subfigure}[b]{0.4\textwidth}
        \centering
        \includegraphics[width=\textwidth]{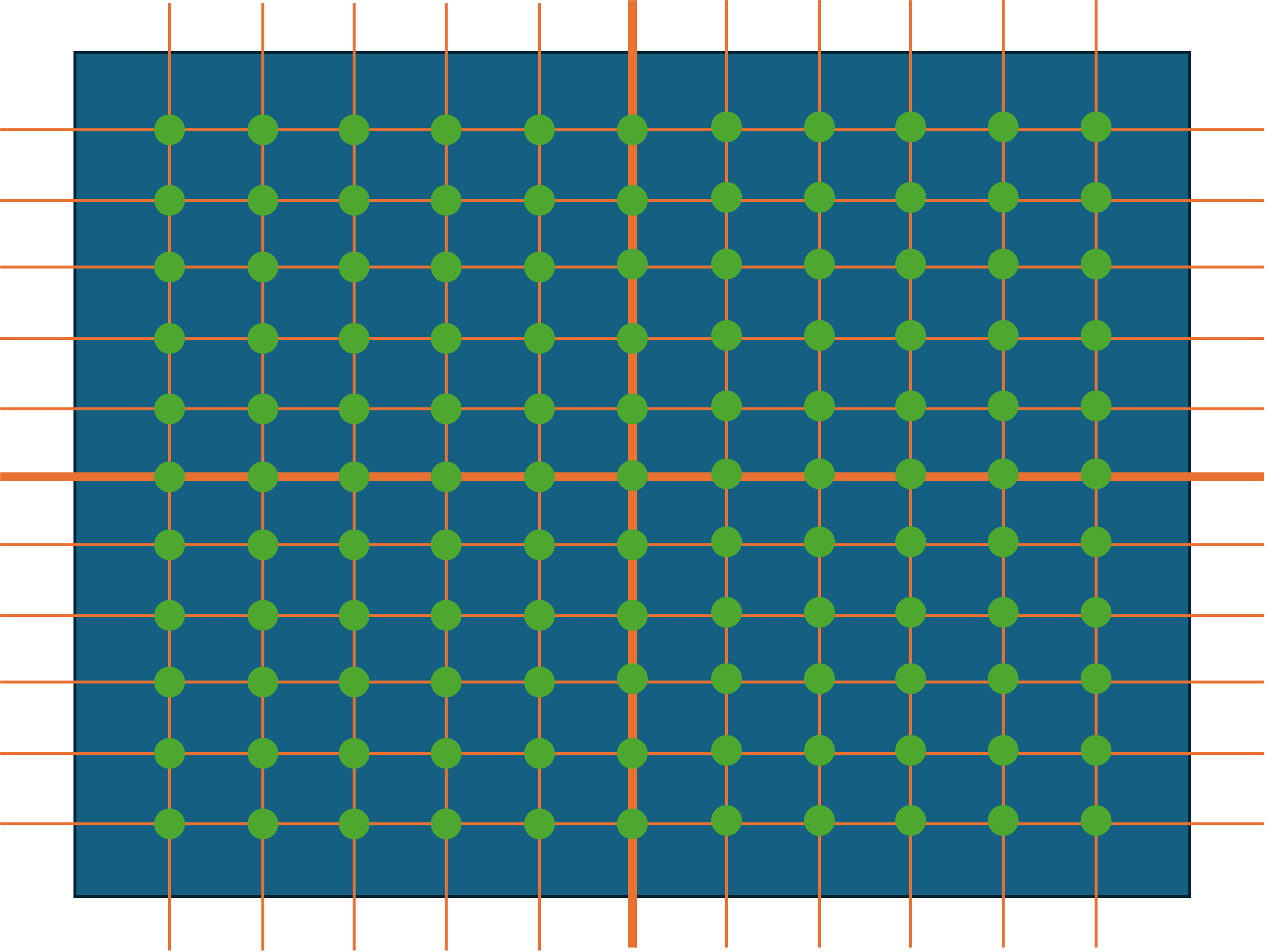}
        \caption{Uniform $10 \times 10$ camera grid.}
    \end{subfigure}
    \hspace{2em}
    \begin{subfigure}[b]{0.4\textwidth}
        \centering
        \includegraphics[width=\textwidth]{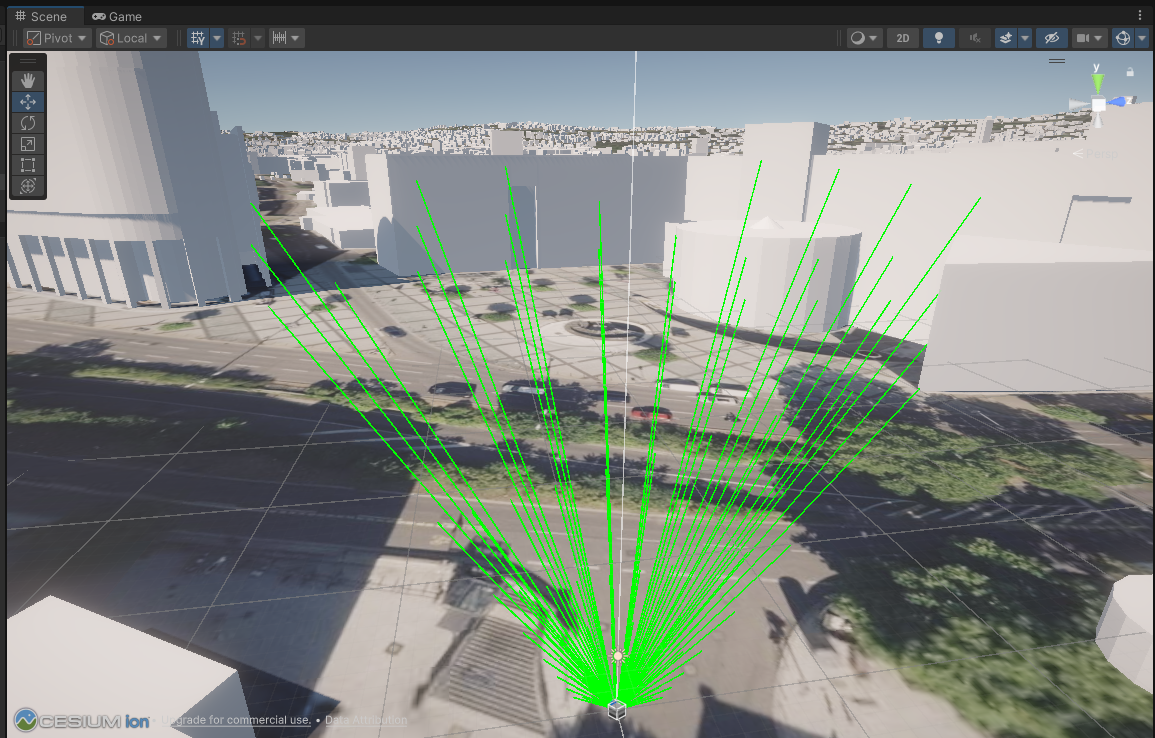}
        \caption{Top-down view of ray casting.}
    \end{subfigure}
    \caption{Overview of the 3D spatial ray-casting strategy within the Digital Twin. The user's dynamic field of view is sampled using (a) a uniform grid of rays to (b) continuously detect intersections with candidate 3D OpenStreetMap building colliders.}
    \label{fig:cameraGrid}
\end{figure}

To identify candidate buildings efficiently, we implemented a uniform spatial ray-casting strategy (Figure \ref{fig:cameraGrid}). The virtual camera replicating the user's Field of View cast a uniform $10 \times 10$ grid of rays (100 rays per frame) up to a distance of 1 km. Because buildings from a street-level perspective occupy substantial visual volume, this sparse sampling strategy reliably captures valid targets while maintaining the low computational overhead necessary for real-time execution. For every ray intersection with a building collider, we extracted the unique OSM Building ID and its semantic metadata (Name, Building Type, Amenity).

Crucially, because head pose is a non-linear estimator for eye gaze \cite{lee_investigating_2018}, we extracted geometric features for every hit, including Euclidean Distance, UV Distance (proximity to the optical center), and Deviation Angle. Furthermore, because visual attention anticipates verbal articulation \cite{griffin_what_2000, stanton_how_2018}, we extracted all visible building data not just during the speech event, but across a 5-second pre-trigger window. This generated a dense, multimodal temporal sequence of candidate buildings for every user query (comprising 2,256 valid ground-truth samples).

\subsection{Subjective Evaluation: Comfort and Workload}
Prior to utilizing this dataset for machine learning, we verified that our collection method elicited natural responses without the confounding effects of simulator discomfort.
Motion sickness was continuously monitored using the Misery Scale (MISC) \cite{MISC2006Bos}. Scores remained exceptionally low, peaking at an average of $1.7 \pm 1.0$ immediately post-VR, but returning to near-zero levels ($0.5 \pm 0.2$) during the Playback sessions, guaranteeing the behavioral integrity of the training samples. Only one participant reported a disproportionately high level of motion sickness, which led to the termination of their session.

\begin{figure}[h]
    \centering
    \begin{subfigure}[b]{0.95\columnwidth}
        \centering
        \includegraphics[width=\textwidth]{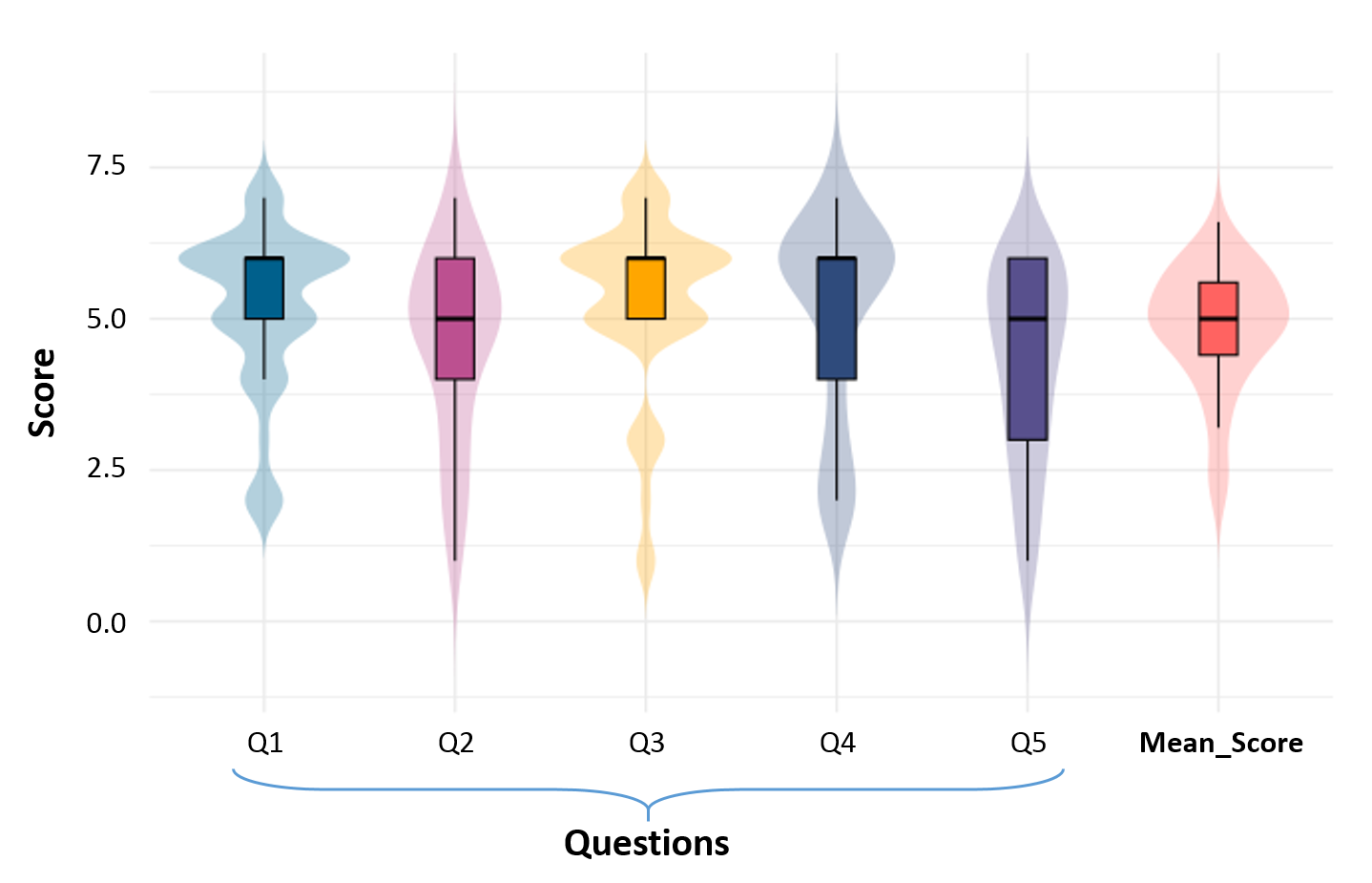}
        \caption{SUS Presence}
    \end{subfigure}
    \hspace{2em}
    \begin{subfigure}[b]{0.95\columnwidth}
        \centering
        \includegraphics[width=\textwidth]{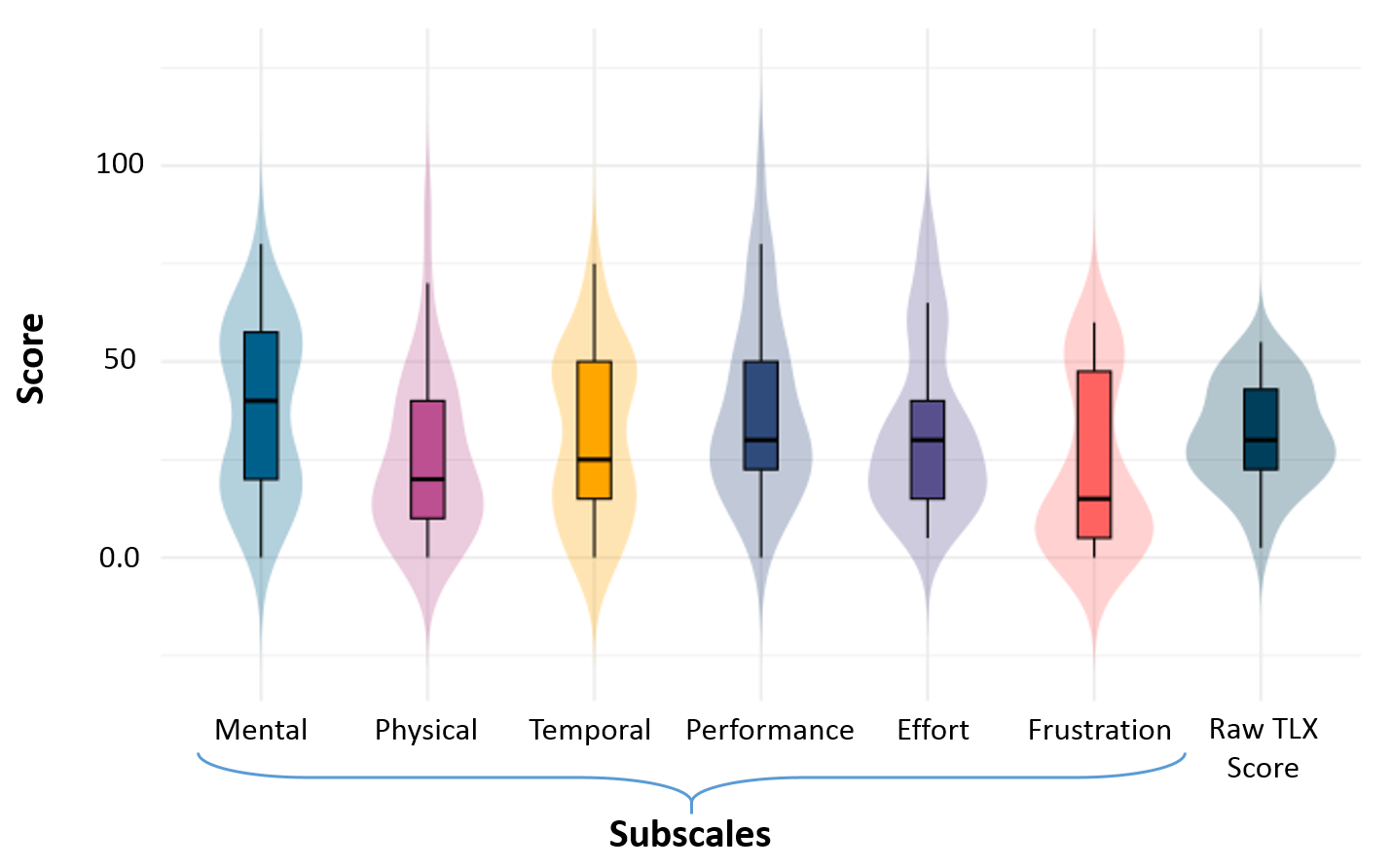}
        \caption{NASA-TLX}
    \end{subfigure}
    \caption{Violin plots of subjective evaluation scores for (a) immersion and (b) cognitive workload.}
    \label{fig:subjective_results}
\end{figure}

Immersion was validated using the Slater-Usoh-Steed (SUS) Presence Questionnaire \cite{slater_depth_1994}. Results indicated strong overall subjective presence (\textbf{Figure \ref{fig:subjective_results}a}). Notably, participants reported exceptionally high scores regarding their sense of "being there" (Q3 $Mdn = 6.0$) and visual realism (Q1 $Mdn = 5.5$), which we heavily attribute to the passive haptics of the physical car seat bridging the digital-physical gap. Finally, the NASA-TLX \cite{hart_nasa-task_2006} indicated a highly suitable interaction paradigm (\textbf{Figure \ref{fig:subjective_results}b}); the overall cognitive workload was low-to-moderate ($32.5 \pm 12.0$ out of 100), characterized by moderate mental demand ($42.0 \pm 15.0$).

\section{Multimodal Fusion Architecture}
\label{sec:architecture}

To process the heterogeneous data streams generated by the Digital Twin and the user study, we designed a custom, multi-branched Transformer network. The architecture is explicitly engineered for real-time inference, utilizing late-stage joint fusion to effectively weigh continuous spatial gaze candidates against the user's discrete spoken intent.

\subsection{Semantic and Spatial Feature Extraction}
The network relies on a dual-branch topology to independently extract intra-modal features before fusion. 

\textbf{User Command Encoding:} The user's transcribed verbal query is processed via pre-trained Sentence-BERT (SBERT)~\cite{reimers2019sbert}. SBERT was strategically selected over standard BERT to generate highly contextual, yet computationally efficient, 384-dimensional sentence embeddings. These embeddings are projected into a compact 64-dimensional latent space using a Feed-Forward Neural Network (FFNN) utilizing Gaussian Error Linear Unit (GELU) activations to maintain smooth gradient flow.

\textbf{Spatial and Contextual Building Encoding:} Parallel to the command processing, the network ingests the temporal sequence of candidate buildings intersected by the ray-casting grid. This branch processes both semantic metadata and geometric telemetry:
\begin{itemize}
    \begin{sloppypar} % withouth sloppypar, the line exceed col width! this is to make the paragraph with relaxed spacing.
    \item \textbf{Semantic Metadata:} Building attributes (Name, Type, Amenity) are individually encoded via SBERT, compressed via FFNNs to 32 dimensions, and concatenated.
    \end{sloppypar}
    \item \textbf{Numerical \& Categorical Telemetry:} Geometric features (distance, UV distance, deviation angle, building height, user head orientation) are normalized. Crucially, the unique OpenStreetMap Building IDs, which are structurally massive integers, undergo a \textit{dense rank normalization} mapping them to a stable $[0, 1]$ range. These normalized IDs are passed through a trainable embedding layer to learn geometric orthogonality, ensuring the model actively separates distinct buildings in the latent space.
    \item \textbf{Temporal Encoding:} To accommodate irregular intersection intervals across the 5-second observation window, raw timestamps are concatenated directly with the numerical attributes before linear projection, acting as a learnable continuous positional embedding.
\end{itemize}

Textual and numerical vectors are fused additively to form a 64-dimensional representation for each building. To model spatial-temporal dependencies between candidates, this sequence is processed through a 4-layer Transformer Encoder ($n_{head}=4$), which dynamically refines features based on the user's evolving navigational context.

\subsection{Cross-Attention Fusion and Decision Head}
To integrate the spatial building sequence with the semantic command, we employ a joint fusion strategy via Multi-Head Cross-Attention~\cite{vaswani2017attention} ($n_{head}=2$). The 64-dimensional SBERT command feature serves as the \textit{Query}, while the Transformer-encoded building sequence serves as both \textit{Keys} and \textit{Values}. 

This mechanism computes attention weights that quantify the semantic relevance of each spatial candidate to the spoken query. To preserve this explicit relevance metric for the final decision, the scalar attention weight is concatenated directly to the building features (yielding a 65-dimensional tensor). Finally, a 4-layer Multi-Layer Perceptron (MLP) compresses this representation ($65 \to 64 \to 32 \to 16 \to 1$) to output raw probabilistic logits for each candidate item.

\subsection{Training Dynamics and Temporal Aggregation}
Training an object referencing model in the context of dynamic urban environment introduces extreme class imbalance; a 5-second interaction window may capture hundreds of negative candidate buildings and only one true positive target. To prevent the network from trivially minimizing loss by predicting the majority negative class, we applied a weighted Binary Cross-Entropy (BCE) objective:

\begin{equation}
\label{eq:bce_function}
\begin{gathered}
\mathcal{L}_{\text{weighted}}(p, y) = \\
-\frac{1}{N} \sum_{i=1}^{N} \Big[ w_{pos} \cdot y_i \cdot \log(p_i) + (1 - y_i) \cdot \log(1 - p_i) \Big]
\end{gathered}
\end{equation}

Where the positive class weight ($w_{pos}$) dynamically scales the penalty for missed targets. Rather than using a fixed global hyperparameter, $w_{pos}$ is computed per-sample based on the exact ratio of negative to positive candidates within that specific sequence ($w_{pos} = N_{neg} / \max(1, N_{pos})$), ensuring stability across varying sequence lengths.

Finally, because the target building typically persists across multiple consecutive frames within the interaction window, the network must resolve temporal redundancy. We implemented a \textbf{temporal sum aggregation} strategy. The raw logits are passed through a sigmoid function to yield probabilities ($P(i)$). These probabilities are aggregated by their unique Building ID, and the system selects the candidate with the maximal cumulative score:

\begin{equation}
\label{eq:prediction}
\hat{y} = \operatorname*{argmax}_{b \in B} \sum_{i \in S_b} P(i)
\end{equation}

\begin{figure*}[h]
    \centering
    \includegraphics[width=1.0\textwidth]{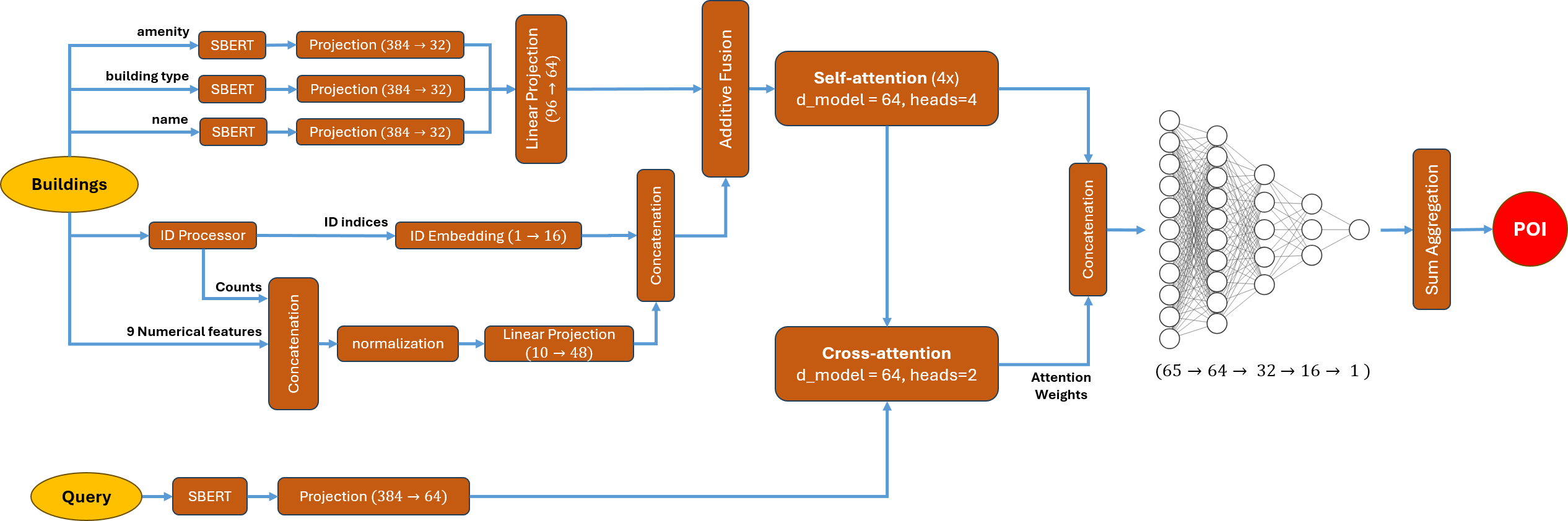}
    \caption{End-to-End Multimodal Inference Pipeline: From heterogeneous spatial and semantic inputs to final POI aggregation.}
    \label{fig:complete_pipeline}
\end{figure*}

Where $B$ is the set of unique buildings and $S_b$ comprises all item indices corresponding to building $b$. By accumulating these temporal confidence scores, the system effectively smooths transient tracking noise and outputs a highly robust, disambiguated Point of Interest prediction (Figure \ref{fig:complete_pipeline}).

\section{Evaluation and Results}
\label{sec:evaluation}

To evaluate the efficacy of our proposed multimodal architecture, we present a quantitative analysis of its predictive performance, confidence margins, and a comprehensive ablation study validating our design choices. 

\subsection{Experimental Setup}
\textbf{Implementation and Hardware:} The model was implemented in PyTorch and trained on a workstation equipped with an Intel Xeon W-2245 CPU, 64 GB RAM, and an NVIDIA Quadro RTX 5000 GPU (16GB VRAM). Utilizing CUDA acceleration, each training epoch averaged 67 seconds. We evaluated the network using standard retrieval and classification metrics calculated on the aggregated output scores, utilizing a 50\% probability decision threshold.

\textbf{Dataset Partitioning:} The initial dataset contained 2,295 samples. After removing 39 incomplete entries due to missing labels, the remaining 2,256 complete samples were partitioned into an 80\% training set, a 10\% validation set, and a 10\% test set using a randomized, sample-level split. We opted for a sample-level split rather than a user-independent one to align with our target in-vehicle scenario, where such an OVR system is repeatedly accessed by a regular group of passengers within the same car. By maximizing the number of participants in the training data, we establish a robust global baseline capable of handling a wide variance of general referencing behaviors. This split strategy aims to challenge the system with distinct audio commands and novel target POIs, while we defer strict cross-participant generalization to future personalization strategies.

\textbf{Training Hyperparameters:}
We trained the model for 100 epochs using a batch size of 4 and the AdamW optimizer ($LR = 10^{-4}$, weight decay $= 10^{-3}$) without a learning rate scheduler. Weights were initialized using PyTorch defaults, excluding the ID Embedding layer which used uniform initialization. Rather than utilizing early stopping, we saved periodic checkpoints and selected the final model post-hoc at epoch 87 based on convergence metrics on the validation set.

\subsection{System Performance}
\begin{figure}[h]
    \centering
    \begin{subfigure}[b]{0.95\columnwidth}
        \centering
        \includegraphics[width=\textwidth]{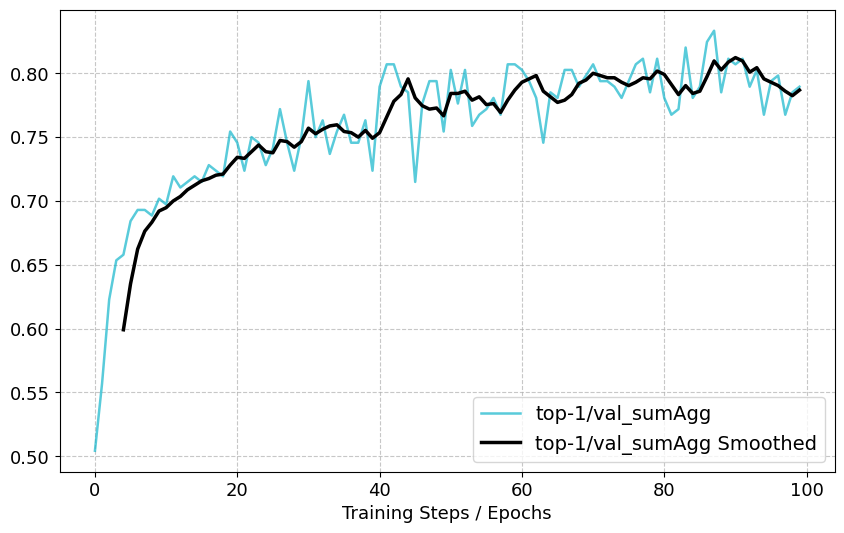}
        \caption{Rank-1 Validation Accuracy}
        \label{fig:val_rank1}
    \end{subfigure}
    \hfill
    \begin{subfigure}[b]{0.95\columnwidth}
        \centering
        \includegraphics[width=\textwidth]{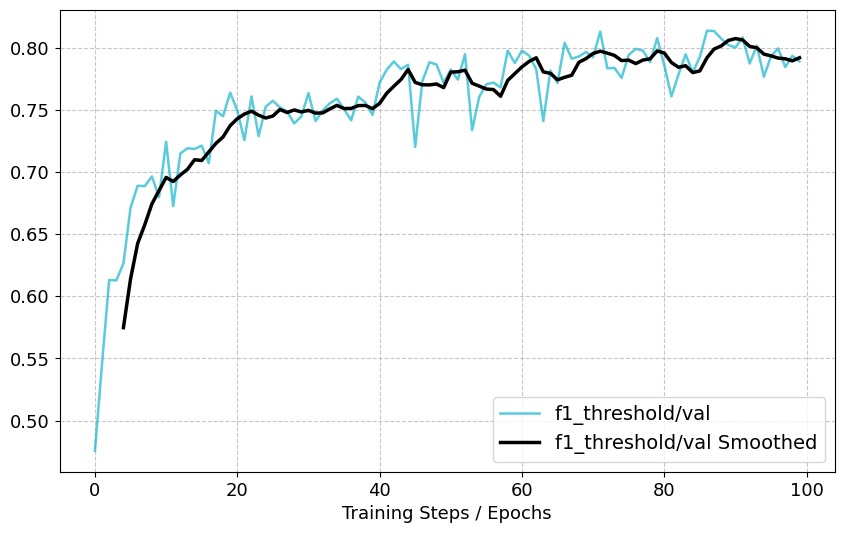}
        \caption{Validation F1-Score}
        \label{fig:val_f1}
    \end{subfigure}
    \caption{Evolution of Rank-1 Accuracy and F1-Score on the validation set. The proposed model achieves asymptotic stability, demonstrating robust generalization and an optimal precision-recall balance.}
    \label{fig:validation_performance}
\end{figure}

The proposed PoI estimation network demonstrated highly robust performance on the validation dataset. As training progressed, the model achieved asymptotic stability, reaching a peak \textbf{Rank-1 Accuracy of 83.33\%} (Figure \ref{fig:val_rank1}). This indicates that in over 83\% of real-world queries, the exact target building was successfully ranked as the absolute top candidate by the system.

To evaluate the model's reliability in ambiguous \textit{near-miss} scenarios, we analyzed the Rank-2 accuracy, which plateaued at 4.39\%. Cumulatively, this yields a \textbf{Top-2 Accuracy of 87.72\%}, proving that even in highly dense urban architectures where multiple buildings fall within the user's spatial gaze cluster, the true target is almost always isolated to the top two candidates.

Regarding classification exactness, the model achieved a Precision of 77.46\% and a Recall of 86.45\%. The harmonic mean of these competing metrics yielded a peak \textbf{F1-Score of 0.813} (Figure \ref{fig:val_f1}). This equilibrium indicates that the network successfully maintains high coverage of true positive spatial targets while strictly minimizing false alarms (the "Midas Touch" problem).

Beyond predictive accuracy, low-latency execution is a strict prerequisite for in-vehicle multimodal interfaces. Timing analysis of the inference phase on the aforementioned workstation with a Quadro RTX 5000 GPU yielded an average latency of $24.33 \pm 11.57$ ms per query (ranging from 12.59 ms to a maximum of 61.46 ms, depending on the number of candidate buildings in the sequence). This computational efficiency translates to an estimated throughput of over 41 samples per second. Because this execution time falls safely below the traditional 100 ms threshold for perceptible system delay in HCI~\cite{card1983psychology}, the proposed architecture successfully guarantees fluid, real-time POI estimation.

\subsection{Confidence and Error Margins}
Beyond binary correctness, we evaluated the robustness of the decision boundaries using margin analysis. The \textit{Correct Winner Margin} (the relative score gap between a correct prediction and the runner-up) showed a continuous positive trend during training, indicating that the model learns to isolate the correct target with high confidence rather than marginal probability. 

Conversely, the \textit{Wrong Winner Margin} (the gap by which an incorrect prediction outperformed the true target) stabilized at approximately 17\%. This relatively low error margin indicates that when the system does fail, the correct target remains a highly competitive "near-miss" rather than being completely ignored by the network.

\subsection{Ablation Study}
To isolate the impact of our specific architectural components and hyperparameter choices, we conducted a comprehensive ablation study (Table \ref{tab:ablation_study}). We evaluated variations in embedding strategies, activation functions, and deep architectural modifications.

\begin{table*}[h]
\centering
\renewcommand{\arraystretch}{1.2}
\begin{tabular}{@{}l|cc@{}}
\toprule
\textbf{Model Configuration} & \textbf{Rank-1 Acc (\%)} & \textbf{F1-Score} \\
\midrule
\textbf{Our Full Architecture} (Trainable IDs, GELU, Sum Agg.) & \textbf{83.33} & \textbf{0.813} \\ 
\midrule
\textit{Unimodal Comparison} & & \\
Vision-Only (Spatial \& Geometric Features) & 76.32 & 0.780 \\ %Epoch 65
\midrule
\textit{Hyperparameter Ablations} & & \\
Fixed Random ID Embeddings & 77.56 & 0.795 \\ % 8-24 Fill in your exact % from the charts here
Leaky ReLU Activation (replacing GELU) & 78.25 & 0.762 \\ % 9-1 Fill in your exact % here
Mean Aggregation (replacing Sum Aggregation) & 80.10 & 0.813 \\ \midrule %9-2-mean Fill in your exact % here
\textit{Architectural Ablations} & & \\
Mod 1: Simplified Numerical Features (No projection) & 79.50 & 0.790 \\ % Fill in your exact % here
Mod 2: Cross-Attention Expansion ($d_{model}=128$) & 79.43 & 0.790 \\ % Fill in your exact % here
Mod 3: Semantic Fusion Only (Late Spatial Concatenation) & 77.56 & 0.770 \\ \bottomrule % Fill in your exact % here
\end{tabular}
\caption{Ablation study comparing our proposed architecture against unimodal, hyperparameter and structural variants. The proposed model consistently outperforms modified configurations.}
\label{tab:ablation_study}
\end{table*}

\textbf{Unimodal Comparison:} To explicitly validate the necessity of our fusion strategy, we evaluated a Vision-Only baseline that relies solely on spatial and geometric gaze features without the user's spoken semantic context. This unimodal configuration achieved a Rank-1 accuracy of 76.32\% and an F1-score of 0.780, representing a performance drop compared to the 83.33\% accuracy of the proposed multimodal baseline. This degradation quantitatively demonstrates the inherent referential ambiguity of purely geometric ray-casting in urban environments. Without the disambiguating power of natural language intent especially in highly clustered areas like city centers, the system struggles to isolate the true target among spatial candidates.

\textbf{Hyperparameter Validation:} The data confirms that utilizing \textit{Trainable ID Embeddings} significantly outperforms fixed random initializations. Allowing the network to iteratively optimize the discrete Building IDs within the continuous latent space effectively learns geometric orthogonality, preventing feature collisions. Furthermore, the GELU activation function slightly outperformed Leaky ReLU, likely due to its smooth differentiability aligning better with the pre-trained SBERT embeddings. Finally, \textit{Sum Aggregation} proved superior to Mean Aggregation, as it better captures the cumulative temporal attention of a user continuously tracking a building across frames.

\textbf{Architectural Necessity:} We tested three structural modifications: simplifying the numerical feature projection (Mod 1), expanding the Cross-Attention capacity to $d_{model}=128$ (Mod 2), and isolating the fusion mechanism strictly to semantic data (user speech and building text attributes), while deferring spatial and geometric features to a late-stage concatenation (Mod 3). The proposed baseline outperformed all three. Notably, the performance degradation observed in Mod 3 reinforces our core hypothesis: semantic speech encoding most effectively helps to resolve referential ambiguity when actively fused with geometric and spatial gaze data, rather than being processed as an isolated stream.
\section{Discussion}
\label{sec:discussion}

\subsection{Advancing State-of-the-Art OVR}
Our multimodal architecture demonstrates a significant leap in OVR capabilities. Prior literature has predominantly approached this challenge under constrained conditions. For instance, Aftab et al.~\cite{aftab_multimodal_2021} achieved 72.2\% accuracy utilizing a stationary vehicle setup with predetermined targets, eliminating the noise of ego-motion. Conversely, Gomaa et al.~\cite{gomaa_looking_2024} tackled dynamic simulator environments but focused on drivers, achieving $\sim$43\% accuracy due to the severe cognitive and physical constraints of operating a vehicle. 

By shifting the paradigm to passengers subjected to continuous visual ego-motion, we leverage a much richer, less constrained behavioral signal. Our system's 83.33\% Rank-1 accuracy validates that fusing an unrestricted, continuous spatial gaze with discrete semantic intent substantially mitigates the referential ambiguity of urban environments. Furthermore, our architectural ablations (Table \ref{tab:ablation_study}) confirm that relying solely on semantic fusion (Mod 3) results in a significant degradation of accuracy. The dual-branch design is strictly necessary to prevent \textit{semantic dominance}, ensuring that the structural and geometric realities of the user's gaze are encoded independently before fusion.

\subsection{Training Dynamics and the Polarization Paradox}
A critical finding during model optimization was the divergence between functional discrimination and probabilistic calibration. After Epoch 20, the model's validation Binary Cross-Entropy (BCE) loss began to rise, even as Rank-1 Accuracy and F1-Scores continued to improve steadily. 

An analysis of the loss landscape revealed this as \textbf{probability polarization}. To minimize entropy in a dataset with extreme class imbalance (hundreds of negative candidate buildings per true positive), the model pushed its predictions toward the absolute extremes of 0 and 1. Due to the logarithmic nature of BCE, exacerbated by our positive class weighting, even a small number of "hard" true positives (assigned low probabilities by the model) caused the aggregate loss to spike disproportionately. 

However, the rising F1-Score confirms that the model's discriminative power was actually improving; it adopted a high-precision strategy, successfully suppressing false positives (the "Midas Touch" problem). Our \textbf{temporal sum aggregation} strategy proved to be the perfect countermeasure to this polarization. By exploiting the natural redundancy of human gaze (passengers fixate on a target repeatedly across frames), the aggregation mechanism allows high-confidence hits to easily overpower localized, polarized false negatives, yielding a highly stable final prediction.

\subsection{Limitations and Real-World Challenges}
While the proposed pipeline successfully operates on continuous-flow urban datasets, deploying OVR in the wild presents distinct physical and data-centric challenges:

\begin{itemize}
    \begin{sloppypar} % withouth sloppypar, the line exceed col width! this is to make the paragraph with relaxed spacing.
    \item \textbf{GNSS Drift and Spatial Alignment:} In dense urban canyons, standard GNSS telemetry suffers from multipath errors. We observed instances where positional drift placed the Digital Twin's virtual camera slightly onto sidewalks or inside adjacent building meshes. This introduces noise into the ray-casting grid. Future deployments require robust Visual-Inertial Odometry integrated tightly with vehicle kinematics to ensure pixel-perfect spatial alignment.
    \end{sloppypar}
    
    \item \textbf{Metadata Sparsity:} While OSM provides excellent geometric collision data, its semantic metadata (amenities, specific business names) is often sparse in residential zones. Our model demonstrated strong resilience to this sparsity by relying heavier on geometric alignment, but integrating commercial geospatial APIs (e.g., Google Maps or Apple Maps) during the ray-casting phase would significantly reduce semantic ambiguity.
    
    \item \textbf{Network Calibration:} To explicitly address the probability polarization observed during training, future iterations of the architecture should replace standard BCE with Focal Loss or incorporate Label Smoothing. These computationally lightweight adjustments would penalize overconfidence and focus gradient updates on hard examples, further stabilizing the loss landscape.
\end{itemize}
\section{Future Work}
\label{sec:future_work}

\begin{sloppypar}
While our multimodal Outside-the-Vehicle Referencing (OVR) framework establishes a robust baseline, future research should explore the following avenues to enhance in-car spatial interaction:
\end{sloppypar}

\begin{itemize}

    \item \textbf{Bridging the Visual-Physical Gap:} While our work successfully isolates and resolves the visual and temporal complexities of continuous-flow real-world transit, it inherently separates the \textit{visually dynamic} experience from the \textit{physically dynamic forces} of live driving. A critical avenue for future work is transitioning this framework into live, on-the-road studies. Evaluating the system under physical vehicular dynamics will allow the community to address hardware resilience against road vibrations and adapt algorithms to the postural shifts caused by active inertial forces.
    
    \item \textbf{Semantic Enrichment and LLMs:} Current reliance on OpenStreetMap limits semantic data in less-mapped areas. Integrating commercial geospatial APIs with Large Language Models (LLMs) would evolve the system into an active conversational agent, capable of processing complex, subjective queries and executing autonomous tasks based directly on the user's view.

    \item \textbf{Personalization via Incremental Learning:} Human pointing behaviors are highly idiosyncratic (e.g., aligning head closely with gaze versus relying on peripheral eye movement) \cite{lee_investigating_2018}. While our current model serves as a robust global baseline capable of handling diverse unseen queries across a broad user population, extending it to a strictly participant-independent setting requires addressing these individual variances. Implementing incremental learning \cite{gomaa_ml-persref_2021, gomaa_looking_2024, reyes_itsallaboutyou_2023} would allow the network to use our current system as a foundational prior, dynamically fine-tuning to a completely new passenger's movement style and refining accuracy by specializing to their specific behavioral traits.
    
    \item \textbf{Eye Gaze Fusion:} While head pose is a robust proxy for user intent, integrating direct eye tracking offers a critical complementary data stream. As automotive XR hardware matures to better handle high-frequency in-car vibrations, fusing eye and head vectors \cite{roider_iseeyourpoint_2018} will resolve edge cases in dense environments by distinguishing between multiple small targets within the same field of view.
    
\end{itemize}
\section{Conclusion}
\label{sec:conclusion}

In this paper, we addressed the challenge of Outside-the-Vehicle Referencing (OVR) for passengers using In-Car Extended Reality (XR). We presented a novel multimodal framework that enables passengers in moving vehicles to seamlessly interact with their external urban environment using natural gaze and speech. To overcome the limitations of stationary setups and the safety challenges of live-traffic testing, we developed a comprehensive data acquisition pipeline. By synchronizing high-fidelity 360-degree urban transit videos with vehicle state data (GNSS positioning, velocity, and orientation), we constructed a high-fidelity geospatial Digital Twin using Cesium and OpenStreetMap.

Leveraging a user study with 46 participants, we captured authentic in-car behavioral patterns and mapped them to semantic building metadata. To resolve the referential ambiguity inherent in dynamic environments, we proposed a lightweight, Transformer-based neural architecture. By utilizing SBERT for semantic encoding and a Cross-Attention mechanism for multimodal fusion, our network effectively weighs continuous spatial gaze candidates against the user's spoken intent. 

Our evaluation demonstrates that this joint fusion approach accurately models the correlation between user gaze and speech in motion. The system achieved a Rank-1 prediction accuracy of 83.33\% and an F1-score of 0.813, successfully placing the true target Point of Interest (POI) within the top-2 predictions in 87.72\% of all evaluated cases. Furthermore, the computational efficiency of the architecture enables real-time inference, making it highly suitable for in-vehicle deployment. By significantly outperforming traditional stationary baselines, this work proves that XR headsets can serve as robust sensing platforms in visually dynamic vehicular contexts, laying a critical foundation for the next generation of intuitive, location-aware automotive infotainment systems.
%\section{Acknowledgment}
\begin{acks}
%\label{sec:acknowledgment}
We extend our deepest gratitude to the 46 participants whose time, effort, and engagement made the user study and data collection possible.    
\end{acks}

\bibliographystyle{ACM-Reference-Format}
\bibliography{content/references}

\appendix
\section{Appendix A: User Study Description}
\label{sec:appendixA}
The informational task sheet provided to each participant during the onboarding phase of our user study is presented below, with minor edits for clarity and academic formatting.

\bigskip
{
\small
\noindent{\bfseries TASK SHEET FOR PARTICIPANTS} 
\smallskip

We aim to understand the interaction and gaze behaviors of passengers when referencing buildings from a moving vehicle. During this study, you will watch 360-degree transit videos in VR. Imagine you are riding in a car with family or friends, and you want to ask a question or draw their attention to a specific building outside. Simply try to comment or ask questions about the buildings in a natural way, just as you would during a real road trip.

\vspace{0.5em}

You may structure your queries and comments in various ways:
\begin{itemize}
    \item Simple queries (no context):
    \begin{itemize}
        \item "What is that building?"
        \item "Wow, look at that building over there!"
    \end{itemize}

    \item Contextual queries (e.g., mentioning business type or features):
    \begin{itemize}
    \item "Look at that red shop! It seems to have great clothes."
    \item "What are the opening hours of this restaurant?"
    \item "What is the price range at this café?"
    \end{itemize}
\end{itemize}

\vspace{0.5em}

Please note the following important study details:

\begin{itemize}
    \item If you are more comfortable speaking German and you can be more productive with it (which can be the case for German participants), please feel free to make your comments and ask questions in German during the VR session.
    \item It is important that you look around and comment on buildings naturally. Please do not feel pressured, and try to avoid artificial behaviors. Think and act as if you are on a normal trip.
    \item When you want to speak about a building, press and hold the designated button on the controller until you finish your sentence and then release it (similar to a walkie-talkie or police radio). We will show you which button to use before we begin.
    \item We will regularly check in with you regarding motion sickness. However, if you feel unwell at any point, please inform us immediately.
    \item When pressing the button to ask a question, please refer to only one specific building at a time and avoid talking about a group of buildings.
    \item For each 12-minute video, we aim to collect around 30 queries (roughly 2 to 4 questions or comments per minute). However, natural behavior is the most important factor. Please do not feel pressured to force queries if it feels unnatural.
\end{itemize}

\section{Appendix B: Detailed Experimental Procedure}
\label{sec:appendixB}

The user study ($N=46$) was conducted in the front passenger seat of a stationary sedan to provide realistic physical constraints. Participants used a tethered Meta Quest 3 HMD (NVIDIA RTX 4080) for 360-degree video playback and a right-hand controller for \textit{push-to-talk} commands. The 75-minute study followed a rigorous sequence:

{
\noindent
\small
\textsf{Onboarding (15 min) $\rightarrow$ VR Session 1 (12 min) $\rightarrow$ Playback 1 (15 min) $\rightarrow$ VR Session 2 (12 min) $\rightarrow$ Playback 2 (15 min) $\rightarrow$ Evaluation ($\sim$6 min)}
}

\begin{enumerate}
    \item \textbf{Onboarding (15 min):} Participants provided informed consent, completed a hardware/environment tutorial, and were instructed to naturally query passing attractions.
    \item \textbf{VR Tasks \& Playback (2 × 27 min):} Participants completed two 12-minute VR sessions, the first featuring a continuous 12-minute route in Stuttgart and the second comprising two 6-minute route segments in Bad Cannstatt and Sindelfingen. The system continuously logged vehicle GNSS, HMD orientation (FoV), and audio (later transcribed/translated via OpenAI Whisper and Gemini). After each VR session, participants exited the vehicle for a 15-minute monitor-based retrospective playback phase. Participants reviewed their recorded POV video alongside the playback of their specific audio commands and the corresponding spatial location on the map. By referencing this synchronized playback, they explicitly identified their intended targets, allowing us to log the exact OpenStreetMap (OSM) Building IDs.
    \item \textbf{Evaluation:} To ensure data integrity, motion sickness (MISC) was monitored continuously during the VR sessions. The SUS (Presence) and NASA-TLX (Workload) questionnaires were completed immediately post-experiment.
\end{enumerate}

}

\end{document}